\documentclass[pra,aps,twocolumn,showpacs,epsfig]{revtex4}
\pdfoutput=1
\usepackage{graphicx,amsmath,amsfonts,amssymb}
\usepackage{latexsym,srcltx}
\usepackage{color}
\begin{document}

\title{Regular and chaotic transport of discrete solitons in asymmetric potentials}
\author{J. Cuevas$^{1}$, B. S\'anchez-Rey$^{1}$, and  M. Salerno$^{2}$}
\affiliation{$^{1}$Departamento de F\'\i sica Aplicada I, E.\ U.\ P.,
Universidad de Sevilla, Virgen de \'Africa 7, 41011 Sevilla,
Spain\\
$^{2}$Dipartimento di Fisica ``E.R. Caianiello'', CNISM  and INFN -
Gruppo Collegato di Salerno,
Universit\`a di Salerno, Via Ponte don Melillo, 84084 Fisciano (SA), Italy}
\begin{abstract}
Ratchet dynamics of topological solitons of the forced and damped
discrete double sine-Gordon system are studied. Directed transport
occurring both in regular and in chaotic regions of the phase space
and its  dependence on damping, amplitude and frequency of the
driving, asymmetry parameter, coupling constant,  has been
extensively investigated. We show that  the passage  from ratchet
phase-locked regime to chaotic ratchets occurs via a period doubling
route to chaos and that, quite surprisingly,  pinned states can
exist inside phase-locking and chaotic transport regions for
intermediate values of the coupling  constant. The possibility to
control chaotic discrete soliton ratchets  by means of both small
subharmonic signals  and  more general periodic drivings, has also
been investigated.
\end{abstract}
\pacs{05.45.Yv, 63.20.Ry, 05.45.-a}
\maketitle

\section{Introduction}
Ratchet dynamics in  nonlinear systems are presently attracting a
great deal of interest due to their relevance in several fields,
including  molecular motors and  Josephson junctions \cite{jap97rmp,r02pr}.
For point particles ratchets the directed
motion achieved in presence of damping and deterministic forces of
zero average, is a direct  consequence of the breaking of the
spatio-temporal symmetries in the system
\cite{fyz00prl,yfzo01epl,dfoyz02pre,sz02pre}. The resulting net motion of
the particle is usually phase-locked to the external driving
\cite{m00prl,bs00pre} and the  deterministic ratchet dynamics
is quite robust to exist also in presence of noise.

Ratchet motion occurring  in  infinite-dimensional continuous
nonlinear  systems with soliton-like solutions (soliton ratchets)
have been extensively investigated in the past years. This has been
done  both in the case of  spatially asymmetric potentials and
single  harmonic driving \cite{m96prl,SQ02} and for symmetric
potentials and  temporarily asymmetric bihharmonic forces
\cite{sz02pre,fzmf02prl,zsc03ijmpb,goldobin01}. Experimental
implementations of soliton ratchets in long Josephson junctions were
also done in  \cite{uckzs04prl} using time-asymmetric biharmonic
currents, and in \cite{cc01prl} and  \cite{goldobin05} using
asymmetric magnetic fields or spatially asymmetric currents.

The mechanism underlying continuous soliton ratchets, also known as
{\it internal mode mechanism},  is  ascribed to the coupling between
the soliton internal degree of freedom ({\it internal mode}) and the
{\it translational mode} of the soliton \cite{SQ02}. This coupling
is induced mainly in the presence of damping and allows an
asymmetric transfer of energy from the external driving to the
internal mode and to the soliton center of mass which generates the
directed motion. Thus, opposite to point particle ratchets for which
the action of the force is always direct on the center of mass, in
soliton ratchets the force acts indirectly on the center of mass and
involves the  vibrational mode of the solitons in an essential
manner (a direct action may exist in the overdamped limit where  the
phenomena becomes negligibly small \cite{m96prl}). This is quite
natural if one think that the soliton is not a rigid object but it
can vibrate while moving,  just like a droplet of liquid does when
dropping. This internal mode mechanism has been confirmed to be of
general validity  for continuous dissipative soliton systems
\cite{SQ02,sz02pre,molinaPRL03,willis04,QSS05,MQSM06}. The
possibility of soliton ratchets in continuous Hamiltonian systems
has been recently demonstrated for the case of the nonlinear
Schr\"odinger equation with asymmetric potentials rapidly flashing
in time \cite{Poletti}.

In contrast to the continuous case, discrete soliton ratchets (DSR)
in nonlinear lattices have been much less investigated. Detailed
studies exist only for ratchet-like behaviors in a dissipative discrete
sine-Gordon chain driven by a  biharmonic force which  breaks the time
symmetry  \cite{ZS06,MC08}, and  for the case of a conservative discrete
nonlinear Schr\"odinger equation describing a 1D array of coupled waveguide optical
resonators \cite{Gorbach}.
In both cases the interplay between discreteness
and nonlinearity introduce  new features in the phenomenon such as:
existence of depinning thresholds for transport, stair-case like
dependence of the mean drift velocity on system parameters,
diffusive and even intermittent ratchet-like dynamics \cite{ZS06}.
The possibility to enhance DSR transport by introducing phase
disorder into the asymmetric biharmonic periodic driving was also
considered \cite{MC08}.

It is interesting, however,  to investigate the phenomenon of DSR in
the case  in which the potential is asymmetric in the field variable
and the system is driven by harmonic forces of zero mean. In this
case the discrete solitons have an  internal degree of freedom even
in absence of damping and driving, a fact which could lead to
qualitative different  behaviors.

The aim of this paper is twofold. From one side, we
extend previous investigations on DSR to the case of asymmetric
potentials. For this we consider the case of a double sine-Gordon
chain with potential function given by
\begin{equation}
    V(x)= C- \frac{1}{4 \pi^2}\cos (2 \pi x)+\frac{\lambda}{8 \pi^2} \sin(4 \pi x), \qquad
    \lambda\in[0,1],
\label{dsine} \end{equation} %
where $\lambda$ is an asymmetry parameter and $C$ is a constant that
fixes the zero of the potential.  Therefore  the system we are going
to consider belongs to the Frenkel-Kontorova family \cite{BK98} and
can be visualized as a chain of
 mechanical  double pendula  connected by a gear of ratio $1/2$ with  nearest-neighbor
 interactions and with the periodic on-site double-sine potential in
(\ref{dsine}) due to the gravity \cite{ms85}.
\begin{figure}[!ht]
\begin{center}
    \includegraphics[width=7cm]{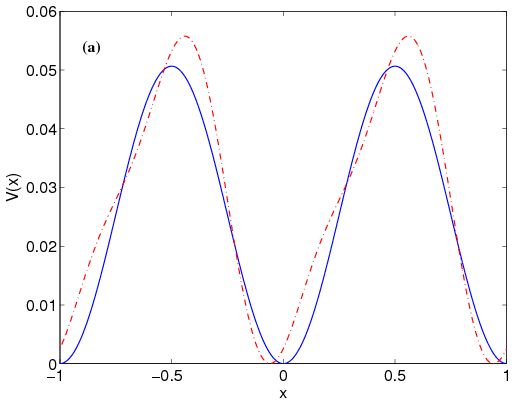}
\end{center}
\begin{center}
    \includegraphics[width=7cm]{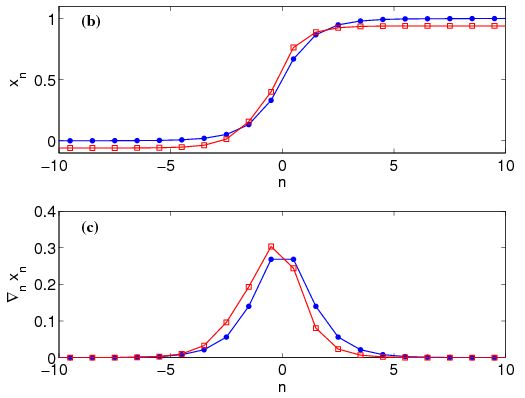}
\caption{(Color online). (a) Comparison between the symmetric potential for
$\lambda=0$ (continuous line) and an asymmetric double sine-Gordon
potential (dashed line) for $\lambda=0.5$. (b) Stationary discrete
kink solutions of the  unperturbed double sine-Gordon chain
corresponding to $\lambda=0$ (circles joined by a continuous line)
and $\lambda=0.5$ (squares joined by a continuous line). The
discrete spatial derivatives of the kink profiles are also shown in
(c) with corresponding symbols. The value of the coupling constant
$\kappa$ is fixed to 1.
 } \label{fig0}
\end{center}
\end{figure}
The dependence of the transport on the damping constant, amplitude
and frequency of the driving force, coupling constant and
asymmetry parameter, is investigated in detail. As a result we show
that in contrast  to  DSR driven by asymmetric forces of zero mean
\cite{ZS06}, kink transport in the present system can be quite
effective not only in correspondence of regular  (phase-locked)
orbits,but also on chaotic trajectories.

>From the other side, we shall investigate the DSR chaotic transport
in some detail both with respect to the onset of chaos and with
respect to the possibility of controlling the chaotic ratchet
trajectories using subharmonic signals which do not break the time
symmetry in the system. To this regard, we introduce a suitable map
in the space of the kink coordinates by means of which we construct
bifurcation patterns. Using this approach, we show that the chaos
onset occurs  via period doubling and that the chaotic region can be
strongly reduced  in favor of periodic phase-locked orbits in
presence of small subharmonic signals. The  complete suppression of
chaos is also investigated by means of elliptic driving functions
with modulus taken as a free parameter to achieve optimal control.
For the regions of parameters we have investigated, we show that the
full chaos suppression can be implemented in any practical context
by means of two properly chosen harmonic signals.

The paper is organized as follows. In Section II we introduce the
model equations and discuss the main properties of the static kink
solutions, including their Peirls-Nabarro barrier and the small
oscillation problems. In section III we investigate the dependence
of the discrete ratchet dynamics, both phase-locked and chaotic, on
the system parameters. In Section IV we concentrate on the ratchet
dynamics associated to chaotic orbits  and on  their stabilization
by means of  small subharmonic modulations of the harmonic driving.
We study the onset of chaos by means of a suitable bifurcation map
and show that chaos occurs via period doubling. Finally, in Section
V we summarize the main conclusions of the paper.
\begin{figure}[!ht]
\begin{center}
\begin{tabular}{cc}
    \includegraphics[width=7cm]{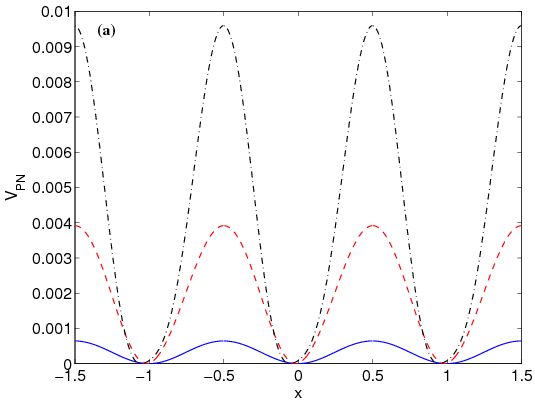}
\end{tabular}
\begin{tabular}{cc}
    \includegraphics[width=7cm]{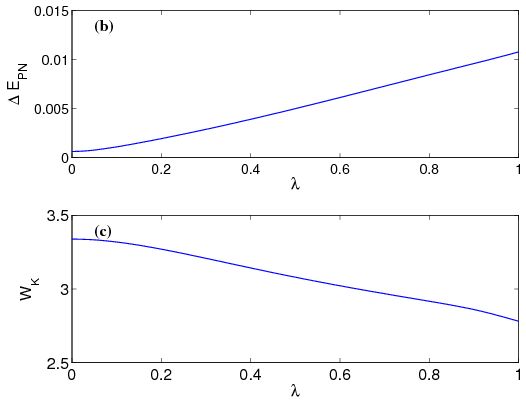}
\end{tabular}
\caption{(Color online). (a) Peierls-Nabarro relief for $\lambda=0$ (continuous
line), $\lambda=0.4$ (dashed line) and $\lambda=0.9$ (dashed-dotted
line). (b) Peierls-Nabarro energy versus lambda. (c) Dependence of
the kink width on the asymmetry parameter.} \label{fig8}
\end{center}
\end{figure}
\section{Model equation}
The chain of double pendula associated to potential (\ref{dsine}) is
described in a dimensionless form by the equation \cite{ms85}
\begin{equation}
    \ddot x_n+\alpha\dot x_n+V'(x_n)+\kappa
    (2x_n-x_{n+1}-x_{n-1}) = \gamma + F(t) + \xi_n (t),
    \label{dynamics}
\end{equation}
with $n\in[-N,N]$, $x_n$ being the displacement of the nth pendulum
with  respect to its equilibrium position, $\gamma$ a constant bias
term,  $\alpha$ the damping coefficient, $\kappa$ the coupling
constant which measures the discreteness of the lattice,
\begin{equation}
    F(t)=\frac{\varepsilon}{2\pi}\cos(\omega t + \theta)
    \label{driver}
\end{equation}
an external symmetric in time periodic driving force of strength $\varepsilon$, frequency
$\omega$ and $\xi_n (t)$ a stochastic term
of zero average and autocorrelation function  fixed as
\begin{equation}
\langle \xi_n (t) \xi_m (t') \rangle =\frac{\alpha D}{\pi} \delta_{nm}
\delta(t-t') \label{noise}
\end{equation}
and modeling the effect of a thermal bath in contact with the
system.  In the following we will investigate mainly deterministic
discrete ratchets in absence of the d.c. bias $\gamma=0$ and in
presence of the a.c. driving of zero average  (\ref{driver}). 
We shall introduce external noise in the system only to check that
the directed motion is associated to single stable attractors, a
fact which  permits to  avoid averaging on initial conditions (e.g.
initial phase $\theta$ of the driving).

In Fig. \ref{fig0}(a) we show the potential chain in (\ref{dsine})
for the case $\lambda=0.5$. Notice that minima of the potential are
located at $n-\delta(\lambda)$,  being
$\delta(\lambda)=\frac{1}{2\pi}
\sin^{-1}\left[\frac{1-\sqrt{1+8\lambda^2}}{4\lambda}\right]$. Since
a discrete kink (anti-kink) can be thought as a string in space
extending from one absolute minimum of the potential to an adjacent
one, we will assume $\lambda \in [0,1]$ to avoid the appearance of
relative minima  in the potential. For large coupling constant
$\kappa$, approximated analytical expressions for kinks and
anti-kinks can be obtained from the corresponding exact solutions of
the continuum limit derived in \cite{SQ02} (notice that for large
$\kappa$ the discrete chain is well approximated by the continuous
double sine-Gordon equation), as
\begin{equation}
\label{soliton} x_n^{\pm} = \delta(\lambda) +\frac{2}{\pi} \tan^{-1}
\left\{\frac{ {\rm sign}(\lambda)\,A\,B}{A-1- \eta \sinh \left[\pm
\frac{\Gamma_n}{2}\sqrt{\frac{AB}{|\lambda|}} \right]} \right\}
\end{equation}
with $A=\sqrt{1 + 8 \lambda^{2}}$, $\eta=2\lambda \sqrt{2(1+A)}$,
$B=\sqrt{2(4\lambda ^{2}-1+A)}$ and
$\Gamma_n=\frac{n-Vt}{\sqrt{1-V^{2}}}$ (the plus and minus signs
refer to the kink and antikink solutions, respectively). For small
values of $\kappa$ the kink solutions must be found numerically. In
the panel (b) of Fig. \ref{fig0} we depict two discrete kinks
obtained numerically for the case of symmetric ($\lambda=0$) and
asymmetric ($\lambda=0.5$) potentials. Notice that in the symmetric
case the kink is  symmetric around the center [this is clearly seen
from the derivative profile shown in panel \ref{fig0}(c)] while in
the case of asymmetric potential the kink is  asymmetric and the
width is a little shrunk.

\begin{figure}[!ht]
\begin{center}
\begin{tabular}{cc}
    \includegraphics[width=7cm]{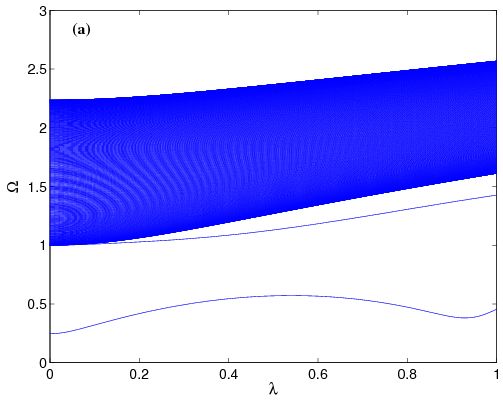}
\end{tabular}
\end{center}
\begin{center}
\begin{tabular}{cc}
    \includegraphics[width=7cm]{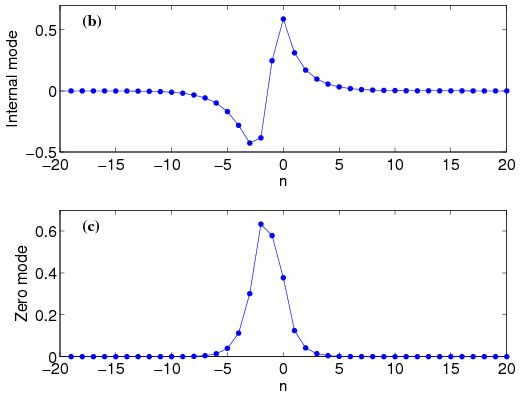}
\end{tabular}
\caption{(Color online). (a) Small oscillations frequency spectrum versus $\lambda$.
The wide band denotes the continuum spectrum, the curve which splits
off from the band corresponds to the internal mode while the bottom
line refers to the translational mode (zero mode). The profiles of
the internal mode and of the translational mode at $\lambda=0.46$
are shown in panels (b) and (c) respectively. The value of the
coupling constant $\kappa$ is fixed to 1.} \label{fig6}
\end{center}
\end{figure}
\subsection{Peierls-Nabarro barrier and depinning}
A main feature of DSR  with respect to continuous soliton ratchets
is the presence in the discrete case of the Peirls-Nabarro  (P-N)
barrier which introduces a depinning threshold for kink transport.
To investigate the dependence of this threshold on the system
parameters we consider static kink solutions of the unperturbed
system solving numerically the nonlinear system of equations
\begin{equation}
    V'(x_n)+\kappa
    (2x_n-x_{n+1}-x_{n-1})=0,
    \qquad n\in[-N,N], \label{unperturbed}
\end{equation}
with the kink boundary condition $x_{N}-x_{-N}=+1$.
\begin{figure}[!ht]
\begin{center}
    \includegraphics[width=7cm]{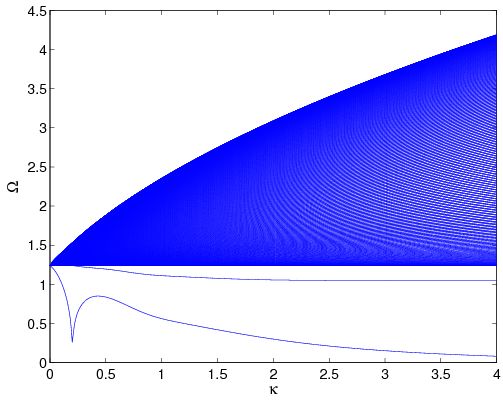}
\end{center}
\caption{(Color online). Small oscillations spectrum versus $\kappa$ for
$\lambda=0.46$. Notice the disappearance of the internal mode
associated with the oscillations of the kink width for $\kappa\le
0.20$} \label{figadd4}
\end{figure}
Starting from a stable kink state we can obtain other (unstable) kink profiles
whose center of mass is shifted with respect to the stable state, and
then  find the dependence of the kink energy on the position of its
center
\begin{equation}
  X_c=\sum_{n=-N}^{n=N} n\, \frac{x_{n+1}-x_{n-1}}{2}
\end{equation}
 In  Fig. \ref{fig8}(a) we show  the resulting Peierls-Nabarro (PN)
relief for three different values of the asymmetry parameter:
$\lambda=0$, $\lambda=0.4$ and $\lambda=0.9$. Note that while for
$\lambda=0$ the PN potential possess the inversion center symmetry
$V_{PN}(X)=V_{PN}(-X)$ this is not true for the case $\lambda \neq
0$, due to the asymmetry of the potential in (\ref{dsine}). In all
cases the P-N potential is a periodic function: $V_{PN}(X+X_0)=
V_{PN}(X)$, with period $X_0$ equal to 1. The difference between the
maxima and minima potential energies, $\Delta E_{PN}$, represents
the activation energy necessary for putting the kink in motion. In
panels (b) and (c) of Fig. \ref{fig8} we have also plotted the
dependence of the activation energy and of the kink width defined
as~\cite{Savin}
\begin{equation}
   W_K=1+2 \sqrt{\sum_{n=-N}^{n=N}(n-X_c)^2 \frac{|x_{n+1}-x_{n-1}|}{2}},
   \label{width}
\end{equation}
on the asymmetry parameter $\lambda$ respectively. One can see that
while the activation energy is an increasing monotonic function of
$\lambda$, the kink width decreases with $\lambda$ in agreement with
what observed for the kink profiles in Fig. \ref{fig0}.
\begin{figure}[!ht]
\begin{center}
\begin{tabular}{cc}
    \includegraphics[width=7cm]{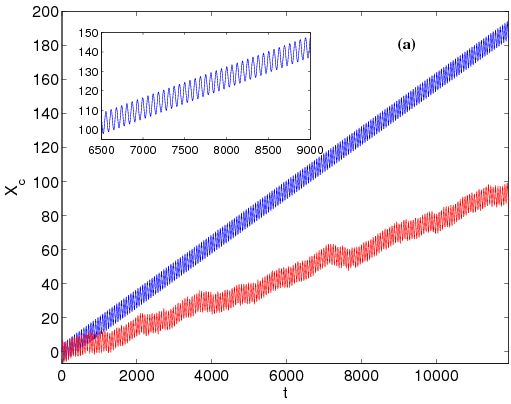}
\end{tabular}
\end{center}
\begin{center}
\begin{tabular}{cc}
    \includegraphics[width=7cm]{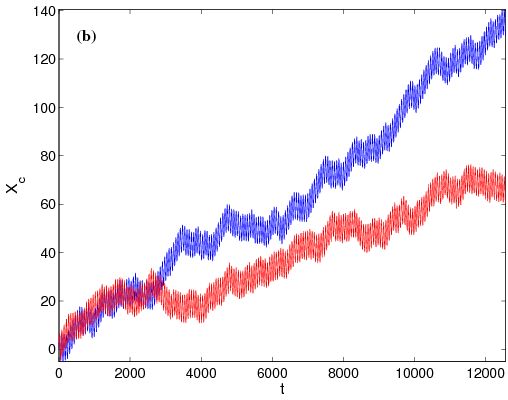}
\end{tabular}
\end{center}
\caption{(Color online). (a) the upper blue line represents phase-locked ratchet
dynamics of the kink center of mass in the absence of noise for
$\lambda=0.46$ and $\varepsilon=0.25$. The inset shows the perfect
synchronization of the motion in a smaller time scale. When some
noise ($D=10^{-4}$) is added the direction of motion is preserved
(bottom red line). (b) Chaotic ratchet dynamics for $\lambda=0.46$
and $\varepsilon=0.24$. Upper blue line corresponds to the
deterministic case while bottom red line corresponds to stochastic
dynamics with $D=10^{-4}$. In all cases, $\alpha=0.1$, $\omega=0.1$
and $\kappa=1$.} \label{fig1}
\end{figure}
It is worth noting that the above symmetry properties of the P-N
potential play important role for  the existence of directed kink
currents in the system. In the lowest approximation, indeed, we have
that the dynamics of the kink in (\ref{soliton}) can be described in
terms of an effective point-particle equation of the type
\cite{sz02pre}
\begin{equation}
\frac{d^2 X(t)}{dt^2} + \alpha \dot{X}(t) + {V'}_{PN}(X) + \tilde F(t) = 0,
\label{cceq}
\end{equation}
where $X(t)$ and $\dot{X}(t)$ are collective coordinates denoting
the position and  the velocity of the kink, respectively, and
$\tilde F(t)$ is an effective driving force proportional to $F(t)$
in Eq. (\ref{driver}). In order to have directed transport for the
kink one must break all symmetries that change sign to the kink
velocity and leave the equation of motion invariant
\cite{dfoyz02pre,sz02pre} (the presence of such symmetries would
imply the presence in the phase space of kink trajectories with
opposite velocities, a fact which is obviously incompatible with
ratchets). Since the P-N potential in our case does not have the
center inversion symmetry and since the force is symmetric in time
the only symmetry which could  link trajectories with opposite
velocities is the one which is indeed an exact symmetry of Eq.
(\ref{cceq}) in absence of damping. In presence of damping, however,
the time inversion symmetry is broken and so are all the symmetries
of Eq. (\ref{cceq})  which link orbits with opposite velocities.
This simple point particle symmetry argument shows how important is
the  damping  in our system to establish directed transport. It
should be remarked, however, that kinks are not exactly point
particles but extended objects with internal structure which also
plays an important role for directed transport, as we shall discuss
in the following.
\begin{figure}[!ht]
\begin{center}
    \includegraphics[width=7cm]{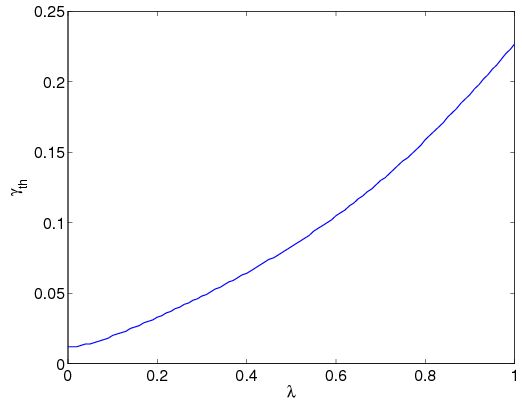}
\caption{(Color online). Depinning threshold for a dc bias $\gamma$ as a function of
the asymmetry parameter $\lambda$. Parameters: $\alpha=0.1$,
$\omega=0.1$, $\varepsilon=0$, $\kappa=1$.} \label{figadd1}
\end{center}
\end{figure}
\subsection{Internal mode of discrete kinks}
As previously mentioned, discrete kinks of the double sine-Gordon
system possess an oscillation of the kink profile (internal mode)
even in the pure Hamiltonian limit. To investigate the existence of
internal modes we have linearized the unperturbed system
(\ref{unperturbed}) around the static kink solution. The numerical
spectrum obtained is shown in  Fig.~\ref{fig6}(a) as a function of
$\lambda$. One can see that, apart the zero mode existing for all
values of  $\lambda$ depicted in panel (c), an internal mode emerges
for $\lambda\ne 0$ below the phonon band. Its shape, plotted in panel
(b), shows a spatial asymmetry induced by the asymmetry of the
potential. This mode plays a crucial role in the appearance of
directed motion through its coupling with the translation of the
kink.

It is worth noting that the existence of that internal mode also depends on
the value of the coupling constant $\kappa$. In Fig.~\ref{figadd4}
it is shown that for a fixed value of $\lambda$ the internal mode
disappears for low enough coupling. In principle one can predict
that no kink net motion must exist below that critical coupling. As
we will see in next section this prediction is confirmed by
simulations corroborating the relevance of the internal mode in the
phenomenon.
\begin{figure}[!ht]
\begin{center}
\begin{tabular}{cc}
    \includegraphics[width=7cm]{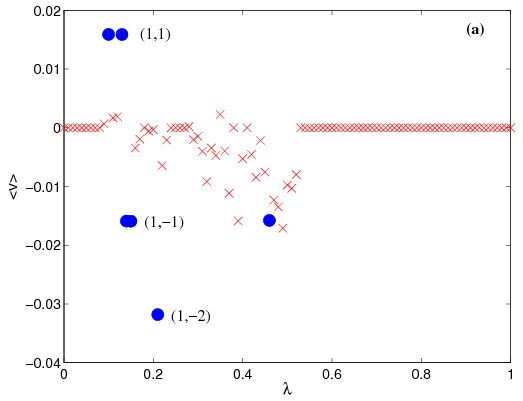}
\end{tabular}
\begin{center}
\end{center}
\begin{tabular}{cc}
    \includegraphics[width=7cm]{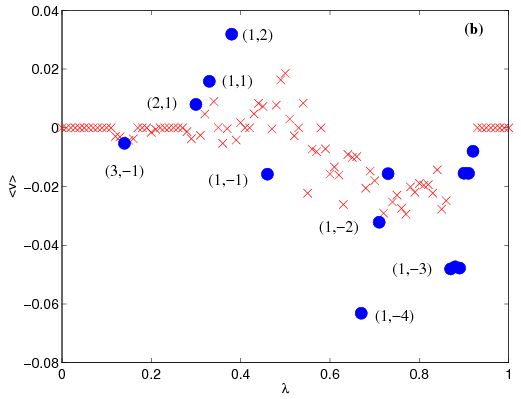}
\end{tabular}
\caption{(Color online). Dependence of $\langle v \rangle$ with respect to $\lambda$
for $\varepsilon=0.1$ (a) and $\varepsilon=0.2$ (b). Full circles
(crosses) correspond to phase-locked (chaotic) dynamics. Numbers in
brackets corresponds to the $(l,k)$ pair of Eq.
(\ref{eq:resonance}). In both panels $\alpha=0.1$, $\omega=0.1$ and
$\kappa=1$.} \label{fig2}
\end{center}
\end{figure}
\section{Phase-locked and chaotic DSR}
As mentioned before,  directed kink motion is expected when
symmetries of Eq.~(\ref{dynamics}) that relate kink solutions with
opposite velocities are broken \cite{dfoyz02pre,sz02pre}. This is
the case for the double-sine Gordon potential (\ref{dsine}) which
violates the symmetry $V(x)=V(-x)$. In order to verify the existence
of kink transport and to study the dependence of the phenomenon on
system parameters, we have performed numerical integration of
Eq.~(\ref{dynamics}) by using a fourth-order Runge-Kutta method with
constant step size $\Delta t=T/200$, being $T$ the period of the
external ac force.  Our system is composed of 400 particles and in
our simulations  static kink was taken as initial condition and
aperiodic boundary conditions $x_N-x_{-N}=1$ were also adopted.

Three main dynamical regimes are observed: i) no net motion; ii)
phase-locked directed motion; iii) chaotic directed motion. Fig.~\ref{fig1} shows
two examples of both kinds of movement. In all cases we checked that
the phenomenon does not depends on the initial phase $\theta$ of the driving
force (\ref{driver}), e.g. it is independent on the choice of the initial conditions so that one
can properly talk about ratchet effect.
\begin{figure}[!ht]
\begin{center}
    \includegraphics[width=7cm]{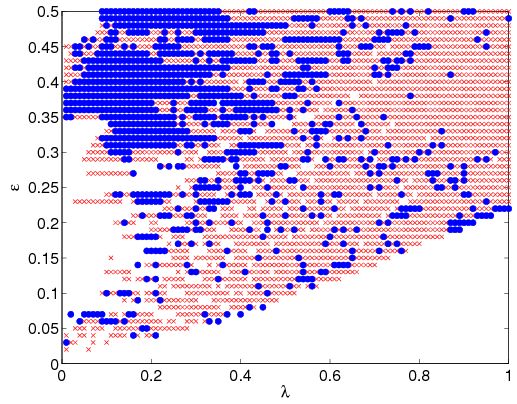}
\caption{(Color online). Dynamical regimes  of the kink motion for different values of
the asymmetry parameter $\lambda$ and the driving amplitude
$\varepsilon$. Full circles (crosses) correspond to phase-locked
(chaotic) dynamics. Blank spaces stand for pinned-kink states.
Parameters: $\alpha=0.1$, $\omega=0.1$, $\kappa=1$.} \label{fig4}
\end{center}
\end{figure}

In the phase-locked dynamics, the kink travels $k$ sites during $l$ periods of the
force, and therefore its velocity is given by
\begin{equation}\label{eq:resonance}
    \langle v \rangle =\frac{k}{l}\frac{\omega}{2\pi}.
\end{equation}
In this regime oscillations on the kink profile are perfectly
synchronized with the motion of the kink center of mass so that the
kink profile is completely reproduced after $l$ periods except for a
shift in space.  This suggests the existence of a coupling between
internal and translational modes of the kink, similarly as for the
continuous case.

In the chaotic regime the kink jumps backward and forward in an
unpredictable manner and the directed motion can only be observed on
a large time scale. In this regime the kink motion is of diffusive
type and resembles the one of a brownian particle. We remark that
the directionality of the  motion in both cases is preserved also in
presence of a noise term of type (\ref{noise}), as one can see from
Fig. \ref{fig1}. This is a consequence of the fact that in the phase
space there is only one attractor corresponding to the net motion of
the kink (in  all investigated cases we have
not found phenomena of hysteresis signaling the presence of more
attractors for the same set of  parameters). This allows to concentrate on the deterministic
phenomenon and compute the mean kink velocity by averaging on time
but not on the initial conditions.

Although in principle  the spatial symmetry of our system is broken
for any value of $\lambda \ne 0$, a distinctive feature of discrete
systems is the existence of a depinning threshold above which the
transport can occur and below which the kink is pinned to a lattice
site. We have used to compute it a dc bias $\gamma$  instead of an
ac force. The result is plotted in Fig.~\ref{figadd1} as a function
of the asymmetry parameter $\lambda$. The depinning threshold
increases monotonously with $\lambda$ in accordance with the rise of
the P-N barrier shown in Fig.~\ref{fig8}.
\begin{figure}[!ht]
\begin{center}
\begin{tabular}{cc}
\includegraphics[width=7cm]{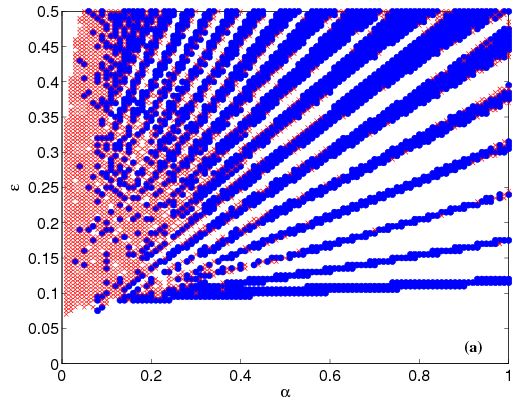}
\end{tabular}
\end{center}
\begin{center}
\begin{tabular}{cc}
\includegraphics[width=7cm]{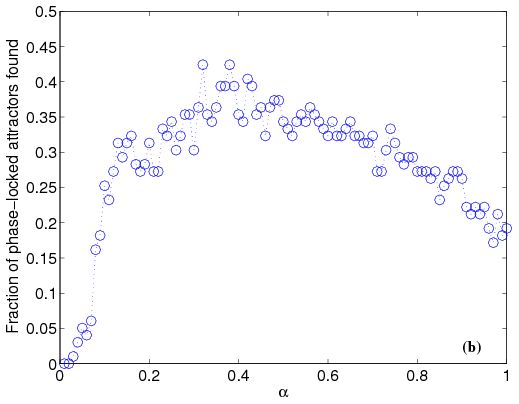}
\end{tabular}
\caption{(Color online). (a) Phase-locked (full circles) and chaotic (crosses)
dynamical regimes of the kink ratchet motion in the
$(\varepsilon,\alpha)$ plane. Blank spaces stand for pinned-kink
states. Other parameters are: $\lambda=0.46$, $\omega=0.1$ and
$\kappa=1$. (b) Fraction of phase-locked attractors found in panel
(a) for each value of the damping.} \label{fig5}
\end{center}
\end{figure}
\begin{figure}[!ht]
\begin{center}
\begin{tabular}{cc}
\includegraphics[width=7.cm]{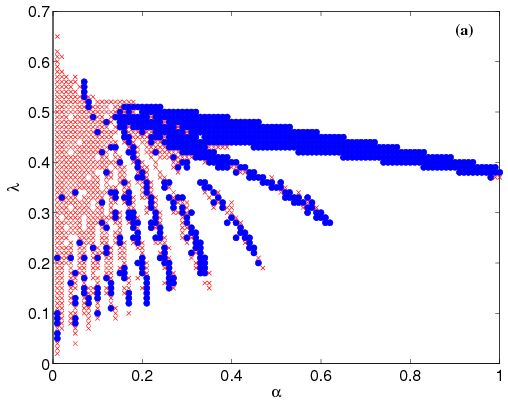}
\end{tabular}
\end{center}
\begin{center}
\begin{tabular}{cc}
\includegraphics[width=7.cm]{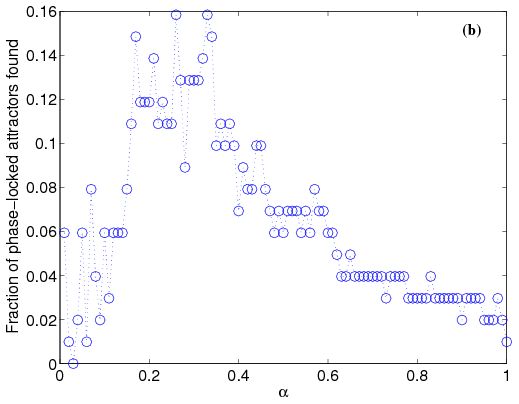}
\end{tabular}
\begin{center}
\begin{tabular}{cc}
\includegraphics[width=7.cm]{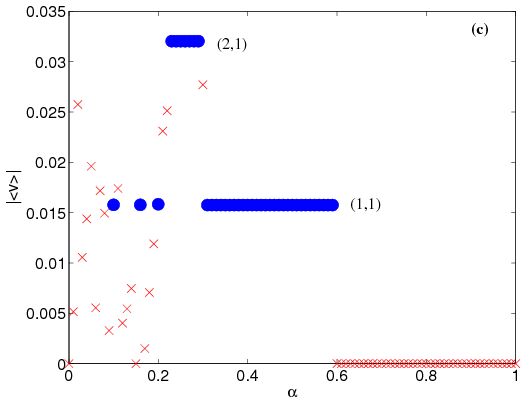}
\end{tabular}
\end{center}
\caption{(Color online). (a) Phase-locked (full circles) and chaotic (crosses)
dynamical regimes of the kink ratchet motion in the
$(\lambda,\alpha)$-plane. Other parameters are:
$\varepsilon=\omega=0.1$, $\kappa=1$. (b) Fraction of phase-locked
attractors found in panel (a) for each value of the damping. (c)
Modulus of the averaged mean drift kink velocity achieved on the
attractors in  panel (a) for $\lambda=0.46$ as a function of the
damping.} \label{fig10}
\end{center}
\end{figure}
For a fixed amplitude $\varepsilon$ of the ac force the dependence
of the mean kink velocity on $\lambda$ is shown in Fig.~\ref{fig2}.
It is interesting to note that even for driving amplitudes above the depinning threshold
a minimal value of the asymmetry parameter is required for the directed motion to start.
By further increasing $\lambda$ a rather complex scenario appears.
Chaotic transport plays a dominant role. Notice that the mean drift
velocity on chaotic orbits can be comparable in some cases with  the
one achieved in the phase-locked regime. Kink dynamics is very
sensitive to parameter values and small changes of them give raise
to sudden changes of the dynamical regime or can even cause a
current inversion. This last finding is actually a striking result
since in the continuous case the sign of $\lambda$ completely
determines the sense of  net motion \cite{QSS05}. Nevertheless, we
have checked that the symmetry property $\langle v(\lambda)
\rangle=-\langle v(-\lambda) \rangle$ existing in the continuous
case is preserved in the case of phase-locked dynamics. The window
of mobility widens of course if we increase $\varepsilon$  and more
resonances appear [see Fig.~\ref{fig2}(b)] but the scenario is
qualitatively the same. In Fig.~\ref{fig4} we have shown the
distribution of regular, chaotic and no transport regimes in the
$(\varepsilon,\lambda)$-plane. We see that whereas the regular kink
transport occurs mainly at small values of $\lambda$ (e.g. in the
region $\lambda<0.3$, $0.3 <\varepsilon <0.5$), chaotic directed
motion becomes dominant at large values of $\lambda$. Notice also
the presence of a large pinning region for small values of
$\lambda$ which extends up  to intermediate values of $\varepsilon$.

The influence on the phenomenon of the damping constant, $\alpha$,
has also been investigated. In  Fig.~\ref{fig5}(a)  we have plotted
a diagram with the phase-locked attractors found on the
($\varepsilon,\alpha)$ plane for fixed $\lambda=0.46$. We have used
in the numerical calculations a step of $0.01$ in the damping
coefficient and a thinner step of $0.005$ in the amplitude of the
driving force. A band structure emerges as $\alpha$ increases and,
significantly, no phase-locked attractors are found when $\alpha
\rightarrow 0$, showing the importance  of damping in the coupling
between the translational an internal modes of the kink
\cite{QSS05,MQSM06}. In panel (b) of Fig.~\ref{fig5} we show the
fraction of phase-locked attractors found for each value of damping;
from it, we can observe a non-monotonous behavior of the curve and
an higher probability of phase locking orbits at intermediate
damping.
\begin{figure}[!ht]
\begin{center}
\begin{tabular}{cc}
\includegraphics[width=7cm]{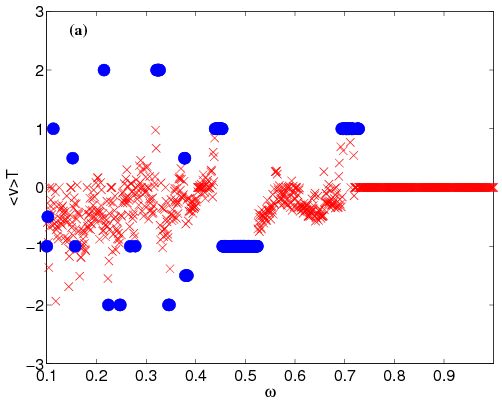}
\end{tabular}
\end{center}
\begin{center}
\begin{tabular}{cc}
\includegraphics[width=7cm]{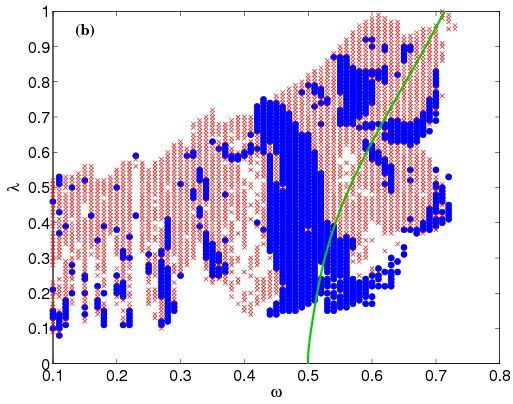}
\end{tabular}
\caption{(Color online). (a) Dependence of the normalized kink velocity $\langle v
\rangle T$ on the driving frequency for $\varepsilon=0.1,
\alpha=0.1, \lambda=0.46$ and $\kappa=1$.  (b) Phase-locked (full
circles) and chaotic transport (crosses)  in the $(\lambda,\omega)$
plane. Blank spaces stand for pinned-kink states while the green
line represents half the frequency of the unperturbed internal mode
at the same value of $\lambda$.
 }
\label{figadd2}
\end{center}
\end{figure}

In  Fig.~\ref{fig10}(a) we report chaotic and phase-locked
attractors found by scanning the ($\lambda, \alpha$)-plane by
keeping the rest of parameters fixed. Again, chaotic behavior is
clearly dominant for low damping while the highest probability to
find a regular attractor occurs at intermediate values of the
damping, as one can see in  panel (b) of Fig.~\ref{fig10}. Since in
the undamped limit the kink mean velocity  must go to zero (due to
the driving force which is symmetric in time), and since for large
damping the internal mode mechanism is practically suppressed (this
strongly reducing the drift velocity), one can expect transport not
only be more probable but also more effective at intermediate values
of the damping. This is indeed what is shown in  panel (c) of
Fig.~\ref{fig10} where the modulus of the averaged mean drift
velocities achieved on the regular and chaotic attractors are shown
as functions of the damping. The fact that at intermediate damping
the coupling between internal vibration and center of mass motion
becomes optimal also correlates with similar studies performed on
continuous \cite{SQ02,QSS05} and discrete \cite{ZS06} soliton
ratchets, as well as,  with  very recent simulations \cite{Bebikhov}
of discrete kink ratchets in an asymmetric potential of polynomial
type free of the Peirls-Nabarro barrier.

The dependence of the mean kink velocity on the driving frequency is
characterized by the alternation of chaos with a series of
phase-locking resonances. This is shown  in Fig.~\ref{figadd2}(a)
where we have depicted the  kink mean velocity normalized to the
driving frequency $\langle v \rangle T= 2\pi \langle v \rangle
/\omega$ so that for phase-locked dynamics $\langle v \rangle T$
coincides with the ratio of rotation numbers $k/l$. We do not find
the monotonic decay observed in the discrete SG model with
biharmonic forces~\cite{ZS06} but, as in that case, kink transport
disappears for high enough frequencies $\omega>0.75$. Notice that
the largest resonant step is achieved around $\omega=0.5$ at the
rotation numbers $k=-1,l=1,$ and that at  low frequencies (e.g.
$\omega < 0.2$) the largest drift velocity is achieved in
correspondence of chaotic trajectories. Also notice in panel (b) of
Fig.~\ref{figadd2} that around $\omega=0.5$ the phase-locked
attractors cover a large region of the $(\lambda-\omega)$-plane
which partially intersects the first subharmonic curve of  the
unperturbed internal mode frequency, this indicating a coupling
between the driving frequency and the internal mode $\Omega_I\approx
2 \omega$.

\begin{figure}[!ht]
\begin{center}
\includegraphics[width=7cm]{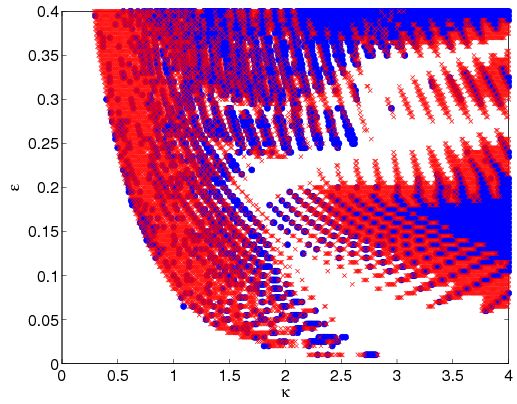}
\caption{(Color online). Phase-locked (full circles) and chaotic transport (crosses)
in the $(\varepsilon,\kappa)$ plane. Blank spaces stand for pinned-kink
states. Parameters are: $\varepsilon=0.1, \alpha=0.1, \omega=0.1,
\lambda=0.46$.} \label{figadd3}
\end{center}
\end{figure}

Lastly,  we have investigated the dependence of the ratchet
phenomenon on the interaction constant $\kappa$. In our model this
parameter measures the discreteness of the system, corresponding
$\kappa \rightarrow \infty$ to the continuum limit. As expected (see
Fig.~\ref{figadd3}) high discreteness prevents propagation of the
kink, and the depinning amplitude of the driving decreases as
$\kappa$ increases. Just above the depinning amplitude chaotic
transport predominates while regular phase-locked motion becomes
more important as $\kappa$ increases. Quite surprisingly, wide
regions in parameter space  in which the kink becomes pinned and the
net motion suppressed appears for intermediate values of the
interaction constant. This feature was not observed in DSR systems
driven by biharmonic forces and indicates that the transport may be
influenced by the local dynamics even if the driving amplitude is
above the depinning threshold for a fixed value of the coupling
constant $\kappa$. This unexpected transport suppression could be
due to anti-resonances with the external driving or to chaos in the
kink motion inside the local Peirls-Nabarro potential which prevents
the escape. The precise mechanism underlying this phenomenon,
however, deserves further investigations.

\section{Chaotic DSR control}
In contrast to continuous soliton ratchet systems which, except for
the case of strongly localized inhomogeneities \cite{Mertens09}, are
very robust against chaos, discrete soliton ratchets exhibit chaos
particularly in the case of asymmetric potential we are dealing
with. We have investigated conditions which allow to reduce the
chaotic orbits in favor of the phase locked dynamics. To this regard
we have added to the sinusoidal driving (\ref{driver}) a weak
subharmonic signal $\eta\cos(\omega t/2)$ in phase with it.

Stabilization effects induced by external weak subharmonic signals
are known to be a general mechanism for suppressing chaos
\cite{Braiman}, both in point-particle systems \cite{Barbi} and in
infinite degree of freedom systems described by partial differential
equations \cite{subharmonic}. The small subharmonic driving is used
not to create new ratchet-like orbits but to stabilize regular
orbits over larger ranges of the system parameters.

\begin{figure}[!ht]
\begin{center}
\begin{tabular}{cc}
    \includegraphics[width=7cm]{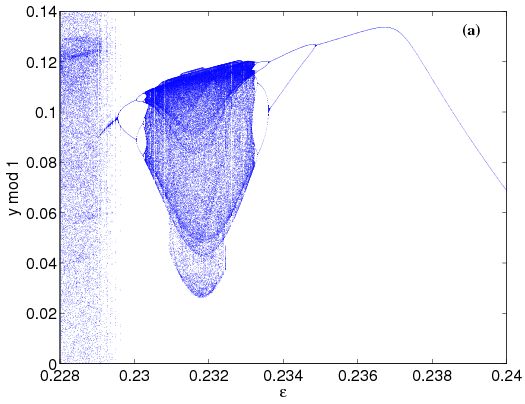}
\end{tabular}
\end{center}
\begin{center}
\begin{tabular}{cc}
    \includegraphics[width=7cm]{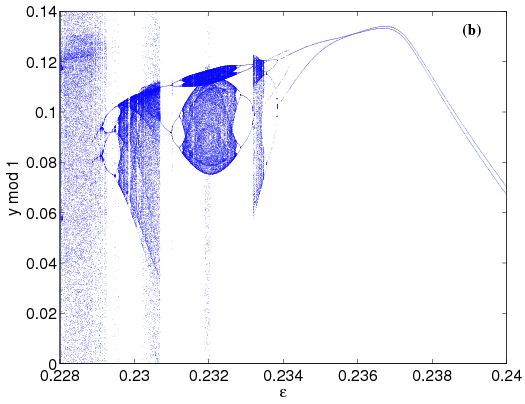}
\end{tabular}
\caption{(Color online). Bifurcation diagrams for $y$ mod 1 with respect to
$\varepsilon$ for $\eta=0$ (a) and $\eta=10^{-4}$ (b). Variable $y$
mod 1 is obtained taken the position of the kink center each period
of the driving after a transient. Other parameters are:
$\lambda=0.4$, $\alpha=0.1$ and $\omega=0.2$.} \label{fig:chaos0}
\end{center}
\end{figure}

In our study we have chosen $\varepsilon$ as bifurcation parameter
since, in principle, in a possible experimental implementation of
the system,  $\varepsilon$ would be easier to control and change
than $\lambda$ for instance.  For simplicity we have focused our
study on an interval of $\varepsilon$ where a destabilization of
phase locking orbits occurs. In  Fig. \ref{fig:chaos0}(a) we have
plotted, for $\eta=0$, a variable $y$ mod 1 versus $\varepsilon$,
where $y$ is defined as the stroboscopic map of $X_c$ with period
$T=2\pi/\omega$ (i.e. $X_c$ taken each period $T$ after a
transient). One can observe phase-locked orbits  represented by
constant values of $y$ mod 1 for a given $\varepsilon$. Transition
to chaos occurs via a cascade of period doubling bifurcations.  In
this graph, for the sake of continuity of the variable $y$, we have
used as initial condition for each step the solution obtained at the
previous one. Another stroboscopic map for the same interval and
$\eta= 10^{-4}$ is shown in  panel (b) of Fig. \ref{fig:chaos0}. We see that
in this case the subharmonic signal, although significantly reduces
the dispersion in $y$ in the chaotic region, it does not give the
full suppression of the chaotic region. This suggests that other
harmonic signals may also be involved in  the full chaos
suppression.

\begin{figure}[!ht]
\begin{center}
\begin{tabular}{cc}
    \includegraphics[width=7cm]{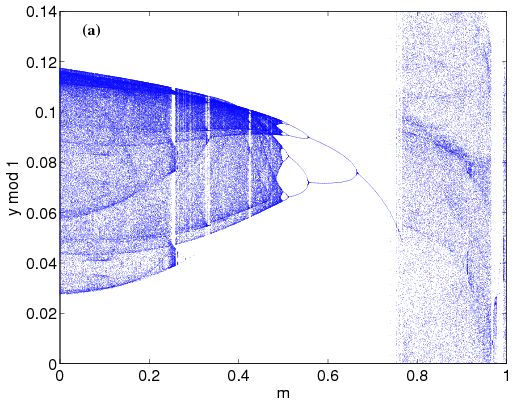}
\end{tabular}
\end{center}
\begin{center}
\begin{tabular}{cc}
    \includegraphics[width=7cm]{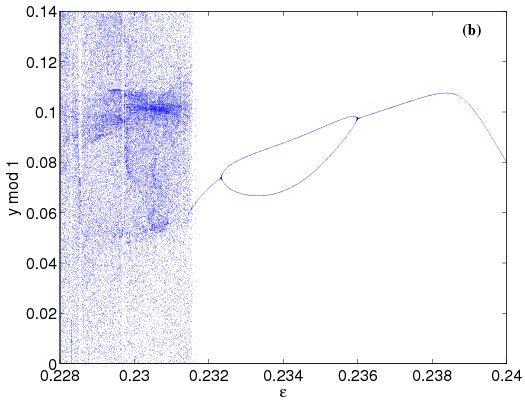}
\end{tabular}
\caption{(Color online). (a) bifurcation diagram for $y$ mod 1 with respect to
parameter $m$ of the elliptic function (\ref{eq:cn})  and
$\varepsilon=0.232$. The amplitude $\varepsilon'$ has been adjusted
according to (\ref{eq:newepsilon}). (b) bifurcation diagram as a
function of $\varepsilon$ for $m=0.7$. In both panels parameters are
the same as in Fig. \ref{fig:chaos0}.} \label{fig:chaos1}
\end{center}
\end{figure}
\begin{figure}[!ht]
\begin{center}
    \includegraphics[width=7cm]{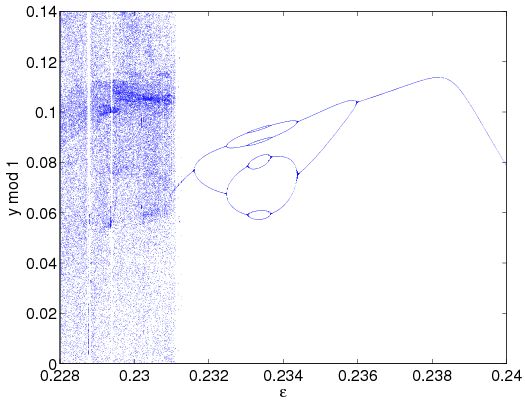}
\caption{(Color online). Bifurcation diagram as a function of
$\varepsilon$ obtained for $m=0.7$ when the elliptic function is
replaced with the first two terms of its Fourier expansion.
Parameters are the same as in Fig. \ref{fig:chaos0}.}
\label{fig:chaos2}
\end{center}
\end{figure}

To find the optimal shape of the small signal for chaos suppression,
it is convenient to use a more general driving in the form of a
Jacobi elliptic function
\begin{equation} \label{eq:cn}
F(t)=\varepsilon' \mbox{cn} \left(\frac{2K}{\pi}\omega t, \,m\right)
\end{equation}
with $\varepsilon'$ the amplitude of the driving, $0\leq m\le 1$ is the
elliptic modulus and $K$ is the complete first order elliptic
integral $K(m)=\int_0^{\pi/2}\,(1-m\sin^2\theta)^{-1/2}\,d\theta$ \cite{Abramowitz}. This
driving allows us to change  the periodicity of the signal
continuously by means of the modulus $m$ used as a free parameter to
optimize the shape. An  elliptic driving, apart from being periodic,
satisfying the symmetry $F(t)=-F(t+T/2)$  and reducing to a
sinusoidal driving (\ref{driver}) in the limit $m=0$, has
the advantage of being expressed in terms of sinusoidal functions as
\begin{equation}
\label{eq:cnexpansion}
F(t)=\frac{2\pi}{m^{1/2}K}
\sum_{n=0}^{\infty} \frac{q^{n+1/2}}{1+q^{2n+1}} \cos [(2n+1)\omega
t]\; ,
\end{equation}
with the coefficient of the series depending on the parameter
$q=\mathrm{e}^{-\pi K(1-m)/K(m)}$  and with  only  odd multiples of
the fundamental frequency entering  the sum. Similar drivings were
also used to suppress chaotic dynamics of Bose-Einstein condensates
loaded into a moving optical lattice \cite{Chacon08}.

The above driving was used to stabilize, for example,  the chaotic
region of our original system around  $\varepsilon=0.232$ in  Fig.
\ref{fig:chaos0}(a). By replacing the sinusoidal subharmonic
driving with the elliptic one (\ref{eq:cn}) and starting from $m=0$,
we have increased $m$ looking for a regularization route to a
phase-locking dynamics. In order to properly compare  with the
sinusoidal case ($m=0$) we have adjusted, while increasing $m$, the
amplitude $\varepsilon'$  so as to keep the first Fourier
coefficient of the expansion (\ref{eq:cnexpansion}) to be
$\varepsilon$, i.e. we performed the regularization by following the
path
\begin{equation}
\label{eq:newepsilon}
  \varepsilon'(m)=\left\{ \begin{array}{ll} \varepsilon & \mbox{if} \;m=0\\
  \varepsilon \frac{m^{1/2}K(1+q)}{2\pi q^{1/2}} &\mbox{if} \; 0 < m < 1  \end{array}
\right. \end{equation} The results are depicted in Fig.
\ref{fig:chaos1}(a). We see that as $m$ is increased a progressive
regularization occurs. The system undergoes an inverse period
doubling sequence of bifurcations until it reaches a periodic orbit
around $m=0.7$. We have chosen this value to compute now a new
stroboscopic map with $\varepsilon$ as bifurcation parameter which
it is shown in  panel (b) of Fig. \ref{fig:chaos1}. Comparison with  Fig.
\ref{fig:chaos0}(a) reveals the complete regularization of the
system for $\varepsilon\gtrsim 0.2316$.

An elliptic driving in practical applications could be difficult  to
implement. We observe, however,  that the main contributions in the
series expansion (\ref{eq:cnexpansion}) come from the first and
third harmonics. Thus, for instance, for $m=0.7$ the coefficients of
the third and fifth harmonics are $0.0802\varepsilon'$ and
$0.00603\varepsilon'$ respectively,
being $\varepsilon'$ the amplitude of the first harmonic.
In Fig. \ref{fig:chaos2} we have depicted the resulting stroboscopic
map when the elliptic function (\ref{eq:cn}) is substituted for the
first two terms of its fourier expansion. Note that the addition of
a third harmonic to the fundamental frequency is indeed enough to
suppress chaos. Therefore a reasonable option in a possible
experimental set-up to control chaos would be to add to the ac
driving a third harmonic conveniently adjusted.

\section{Conclusions}
We have studied the ratchet phenomena induced by asymmetric
potentials on discrete topological solitons (kinks and antikinks) of
the discrete damped double sine-Gordon equation in the presence of
an harmonic driving. The dependence of the mean velocity of the kink
on the asymmetry parameter, damping, amplitude of the driving,
frequency of the driving and on the coupling constant, has been
extensively investigated. We have shown that in contrast with DSR
driven by asymmetric forces of zero mean \cite{ZS06}, kink transport
can be quite effective not only on regular (phase-locking) orbits,
as it is usual for soliton ratchets \cite{SQ02}, but also on chaotic
trajectories. In particular, we have showed that although the mean
kink velocity in the chaotic region is typically smaller than the
one achieved on phase-locked orbits, there are parameter values for
which the chaotic transport becomes more efficient than the one in
the phase locking regime (this case is  observed, for example, in
Fig. \ref{figadd2}(a) for  small values of the driving frequencies,
e.g. for $\omega < 0.2$).

Our investigation of the dependence of the kink transport on the
damping constant shows that a phase locking regime and therefore an
efficient transport is more probable at intermediate values of
damping. This fact correlates with previous investigations and
explanations in terms of the internal mode mechanism.

The mechanism of the kink transport also depends strongly on the
coupling constant. We find that for large values of the coupling
constant the kink transport is dominated by the regular phase locked
motion while the diffusive chaotic transport becomes important for
small values of $\kappa$. Quite surprisingly, we have also found
that for intermediate values of $\kappa$ a wide region of pinned
kink states can exist.

The ratchet dynamics in the chaotic regime has been investigated in
terms of a reduced map for the center of the kink which allows us to
show that the onset of chaos occurs via period doubling
bifurcations. The possibility to control chaotic ratchet dynamics
has also been investigated by means of an additional sub-harmonic
driving which does not break the time symmetry of the system. In
particular we have shown that the chaotic region can be strongly
reduced  in favor of periodic phase-locked orbits even for very
small strengths of the subharmonic signal in comparison with the
fundamental driving. To fully  suppress chaos we have used a more
general driving in the form of an elliptic function with the modulus
taken as a free parameter to modulate the signal and to find the
optimal regularization route for the system dynamics. For the region
of parameters investigated, we have found that in any practical
context this would lead to phase-locking ratchet transport and to
chaos suppression simply by adding a third harmonic conveniently
adjusted to the fundamental driving.

Finally, we remark that the features of DSR investigated in this paper may
be of general validity also for other systems undergoing structural changes
which induce an asymmetry  in the potential energy.

\subsection*{Acknowledgments}
 We acknowledge F. Palmero for helpful comments. JC and BS-R acknowledge
 financial support from the MICINN project FIS2008-04848. BS-R also
 acknowledges financial support from the MICINN project FIS2008-02873.
MS acknowledges partial support from a MIUR initiative about research
 projects of national relevance (PRIN-2008).

\end{document}